# Participant-spectator matter at the energy of vanishing flow

S. Gautam, Aman D. Sood*, and Rajeev K. Puri
*Physics Department, Panjab University, Chandigarh-160014, INDIA*
*email: amandsood@gmail.com*

## Introduction

One of the alluring goals of heavy ion reactions at intermediate energies is to study the nuclear properties of hot and dense matter. Information about equation of state (EOS), in medium nucleon-nucleon cross-section [1-2], incident energy, and impact parameter [3], all play a major role. The nuclear interactions are attractive at low incident energies allowing the scattering of nucleons at negative deflection angle. These interactions, however, turn repulsive, once the incident energy is high enough. In this energy variation, these interactions counterbalance each other, at a particular point. This leadls to no net preference to the nuclear transverse in-plane flow, therefore, net flow disappears [4]. The energy at which net flow disappears is called energy of vanishing flow (EVF).

Recently, in Ref. [5], for the first time, we reported that the participant-spectator matter at EVF is quite insensitive to the mass of the colliding system. It, therefore, can act as a barometer for the study of EVF. Here, we plan to extend the study of participant-spectator matter at EVF for other equations of state, impact parameter as well as for momentum dependent interactions (MDI). We plan to look whether the above participant-spectator matter demonstration still holds well or not.

## Model

The present study is carried out within the framework of quantum molecular dynamics (QMD) model [1]. In this model, nucleon are propagated using classical equations of motion;

$$\dot{r}_i = \frac{\partial H}{\partial p_i}; \quad \dot{p}_i = -\frac{\partial H}{\partial r_i}, \quad (1)$$

where H is the Hamiltonian and is given by

$$H = \sum_i^A \frac{p_i^2}{2m_i} + \sum_i^A (V_i^{Skyrme} + V_i^{Yuk} + V_i^{Coul} + V_i^{mdi}), \quad (2)$$

where $V_i^{Skyrme}$, $V_i^{Yuk}$, $V_i^{Coul}$ and $V_i^{mdi}$ are respectively, the Skyrme, Yukawa, Coulomb, and MDI potentials. The MDI is obtained by parameterizing the momentum dependence of the real part of the optical potential. The final form of the potential reads as:

$$V^{MDI} = t_4 \ln^2(t_5(\vec{p}_1 - \vec{p}_2)^2 + 1)\delta(\vec{r}_1 - \vec{r}_2). \quad (3)$$

Here $t_4$ = 1.57 MeV and $t_5 = 5 \times 10^{-4}$ MeV$^{-2}$.

## Results and discussion

The present study is carried out for symmetric systems; $^{12}C+^{12}C$, $^{20}Ne+^{20}Ne$, $^{40}Ca+^{40}Ca$, $^{58}Ni+^{58}Ni$, $^{93}Nb+^{93}Nb$, $^{131}Xe+^{131}Xe$, $^{197}Au+^{197}Au$ and $^{238}U+^{238}U$. The analysis is done with soft and hard EOS, with and without MDI using σ=55mb. The hard (soft) equation of state has been belled as Hard (Soft). Whereas the hard equation of state with MDI has been labeled as HMD. The simulations are carried out at small energy steps between 30 MeV/A and 250 MeV/A.

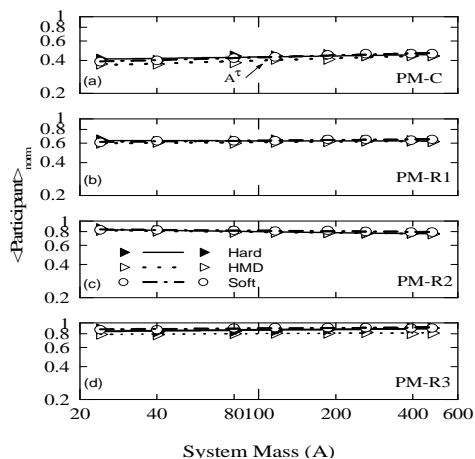

**Fig. 1** The participant matter as a function of system size at their corresponding EVF. (a) is for PM-C. (b), (c), and (d) are for PM-R1, PM-R2, and PM-R3 respectively.

The normalised participant-spectator matter is extracted with two different procedures [5]. (a) All nucleons having experienced at least one collision are counted as participant matter (labeled as PM-C). The remaining matter is labelled as spectator matter (labeled as SM-C). These definitions however, are more of a theoretical interest since the matter defined in these zones can not be measured. Alternatively, we can define participant-spectator matter in terms of the rapidity distribution:

$$Y(i) = \frac{1}{2}\ln\frac{\mathbf{E}(i)+\mathbf{p}_z(i)}{\mathbf{E}(i)-\mathbf{P}_z(i)} \quad (4)$$

where $\mathbf{E}(i)$ and $\mathbf{p_z}(i)$ are respectively, the total energy and longitudinal momentum of the $i^{th}$ particle. We shall rather use a reduced rapidity $Y_{red} = Y(i)/Y_{beam}(i)$. Here we define normalized participant matter by imposing three different cuts in the rapidity distribution: (i) all nucleons with $-1.0 \leq Y_{red}(i) \leq +1.0$ (marked as PM-R1), (ii) $-0.75 \leq Y_{red}(i) \leq +0.75$ (marked as PM-R2), and (iii) $-0.5 \leq Y_{red}(i) \leq +0.5$ (marked as PM-R3).

In fig. 1(a-d), we display the normalised participant matter as a function of combined mass of the system at their corresponding EVF. All the reactions are at semicentral geometry of $b/b_{max} = 0.4$. From fig., it is clear that participant matter (defined in terms of different rapidity distribution cuts) is nearly same for Hard, HMD and Soft EOS, thus indicating that at EVF, the sensitivity of participant matter is insignificant towards nuclear matter EOS.

In fig. 2, the normalised participant matter is shown as a function of system mass for different colliding geometries. All types of participant matter is found to be nearly independent of colliding geometries.

From the above discussion, the following conclusions are visible. The participant matter shows almost mass independent behaviour irrespective of model parameters for central and semicentral reactions. For a given set of model parameters, EVF is higher for lighter colliding nuclei leading to higher density. This results in frequent nn collisions resulting in mass independent behaviour of participant matter.

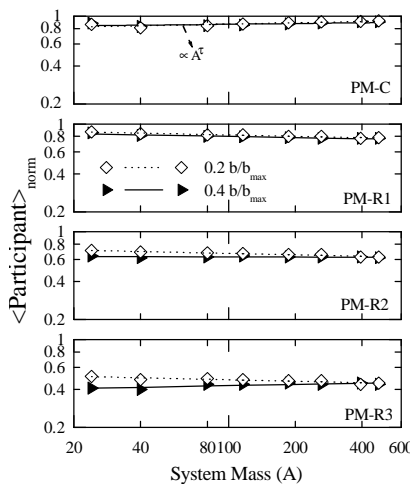

**Fig. 2** The participant matter as a function of system size for different colliding geometeries. The open diamonds are for $b/b_{max} = 0.2$ whereas solid triangles are for $b/b_{max} = 0.4$.